\documentclass[aip,amsmath,amssymb,reprint,]{revtex4-1}
\usepackage{graphicx,xcolor,bm}
\usepackage{etoolbox}

\makeatletter
\def\@email#1#2{%
 \endgroup
 \patchcmd{\titleblock@produce}
  {\frontmatter@RRAPformat}
  {\frontmatter@RRAPformat{\produce@RRAP{*#1\href{mailto:#2}{#2}}}\frontmatter@RRAPformat}
  {}{}
}%
\makeatother
\begin{document}

\title[Magnetocaloric Effect of  Pure and Diluted  Quantum Magnet Yb$_3$Ga$_5$O$_{12}$]
{Magnetocaloric Effect of  Pure and Diluted  Quantum Magnet Yb$_3$Ga$_5$O$_{12}$}
\author{E. Riordan}
\affiliation{Institut N\'eel, CNRS, F-38000 Grenoble, France}
\author{E.Lhotel}
\email{elsa.lhotel@neel.cnrs.fr}
\affiliation{Institut N\'eel, CNRS, F-38000 Grenoble, France}
\author{N.-R. Camara}
\affiliation{Univ. Grenoble Alpes, CEA, IRIG, PHELIQS, F-38000 Grenoble, France}
\author{C. Marin}%
\affiliation{Univ. Grenoble Alpes, CEA, IRIG, PHELIQS, F-38000 Grenoble, France}
\author{M. E. Zhitomirsky}%
\email{mike.zhitomirsky@cea.fr}
\affiliation{Univ. Grenoble Alpes, CEA, IRIG, PHELIQS, F-38000 Grenoble, France}

\date{\today}

\begin{abstract}
The magnetocaloric effect  in the quantum dipolar magnet Yb$_3$Ga$_5$O$_{12}$ is studied both for pure material and with non-magnetic substitution: (Yb$_{1-x}$Y$_x$)$_3$Ga$_5$O$_{12}$. Magnetization measurements have been performed on a  single crystal, $x=0$, and on powder samples with $x = 0.2$ and 0.4 in the temperature range between 70 mK to 300 K and in magnetic fields up to 8 T. The magnetic entropy change 
$\Delta S_m$, a key figure of merit for adiabatic demagnetization refrigeration, has been derived from the magnetization data. The $x=0.2$ sample exhibits the volumetric entropy variation comparable to, and at low fields even enhanced relative to, the pure compound. In contrast, the 40\%\ diluted sample shows a reduced effect, consistent with 
the conventional dilution picture. The Curie-Weiss law fits reveal positive Curie temperatures in both diluted samples, indicating the persistence of ferromagnetic correlations. The robustness of the magnetocaloric response upon moderate dilution highlights the potential of YbGG-based materials for low-temperature magnetic cooling applications, particularly in addressing thermal conductivity challenges through the chemical substitution without compromising cooling power.
\end{abstract}

\maketitle

The adiabatic demagnetization refrigeration (ADR)  is an efficient and  practical method to reach  low temperatures, especially for the situations where  the absence of cryogenic liquids is desirable.\cite{Shirron2014}  
An example of this is the space industry, where magnetic refrigeration is an ideal technique for cooling the telescope detectors on satellites. Additionally, the ever-rising prices of $^3$He and $^4$He provide further motivation for the ADR applications below 4~K. \cite{Cho2009,Kramer2019} To achieve a high cooling power the refrigerant materials must exhibit a large magnetocaloric effect in the operational temperature range. \cite{Wikus2014}
The common magnetic refrigerants are paramagnetic salts, which are not necessarily optimal in terms of the cooling power density as their zero-field magnetic entropy measured per volume is relatively low. Magnetically frustrated materials provide another class  of promising refrigerants for low-temperature ADR applications owing to suppression of magnetic order and the presence of a macroscopic number of low-energy excitation modes. \cite{Zhitomirsky2003} 

The dipolar magnet Yb$_3$Ga$_5$O$_{12}$ (YbGG) has previously been identified as a  promising refrigerant material. \cite{Paixao2020} In YbGG, magnetic Yb$^{3+}$ ions occupy three-dimensional networks of corner-sharing triangles, forming two equivalent interpenetrating sublattices. In the literature, such networks are known as hyperkagome lattices. Long-range dipolar and nearest-neighbour antiferromagnetic exchange interactions on this lattice lead to a frustrated magnetic state above the magnetic transition at $T_{\rm c} = 54$~mK. \cite{Filippi1980,Raymond2024} In this article, we report the magnetocaloric properties of YbGG above the transition temperature. We show that these properties are robust with respect to dilution on the magnetic site and can even be enhanced.

In our study, we have used  three samples: an undiluted single crystal (parallelepiped with dimensions $0.93 \times 0.95 \times 4.37$ mm$^3$) and two powder samples with 20\% and 40\% dilution of Yb by Y ions. 
The  single crystals were grown by the floating zone method using an optical furnace. Feed rods have been
prepared by the solid-state reaction from the high-purity starting powders of Yb$_2$O$_3$ (99.99\%) and 
Ga$_2$O$_3$ (99.99\%). Isostatic pressing (250 MPa) and high temperature sintering up to 1400$^\circ$C are essential to obtain high-density ceramic rods (length $\sim 100$~mm). The molten zone has been stabilized through a flow of argon under pressure (0.5 MPa) at a growth rate of 10 mm per hour. Typical size of YbGG  rods were 8 mm of diameter and  several cm in length. Single crystallinity and phase purity were checked by the X-ray Laue technique and the X-ray powder diffraction. The growth direction is found to be along the $[1 0 0]$ crystallographic axis. 
The X-ray diffraction (XRD) analysis confirmed that both the 20\% and 40\% substituted samples crystallized in the pure cubic garnet structure (space group $Ia\bar{3}d$), with no detectable secondary phase. The diffraction patterns yielded lattice parameters consistent with those of the undoped YbGG, indicating successful incorporation of Y$^{3+}$ ions into the garnet lattice.

 Magnetization measurements were performed in the 70~mK--300 K range. A commercial VSM-SQUID Quantum Design PPMS was used in the 4-300 K range and a SQUID magnetometer developed at the Institut N\'eel and equipped with a dilution refrigerator was used for measurements below 4.2 K \cite{Paulsen2001}. Magnetization measurements on the pure YbGG samples have been performed with the magnetic field applied parallel to the crystallographic [100] axis. This axis is the longest axis of the sample and the demagnetization factor along this direction is estimated to be  $D=0.0947$ (SI units), all measurements of the pure sample have been corrected for demagnetization effects.
 For the diluted powder samples, the sample is mixed with vacuum grease and wrapped in a copper pouch to ensure good thermal contact. 
 The pouches containing the samples are approximately flat squares in shape and positioned such that the magnetic field is parallel to the plane of the square. Accurately estimating the demagnetization factor for these samples would be very difficult, so the corresponding data are presented here without correction but we expect the factor to be small (certainly under $D=1/3$). It is important to note that demagnetization effects reduce the internal magnetic field that the sample ``sees", e.g., for the pure sample the field is reduced by up to 0.025~T, this can be important to bear in mind when comparing between samples of different shapes for the same applied magnetic field. 


\begin{figure}[t]
\includegraphics[width=\linewidth,keepaspectratio]{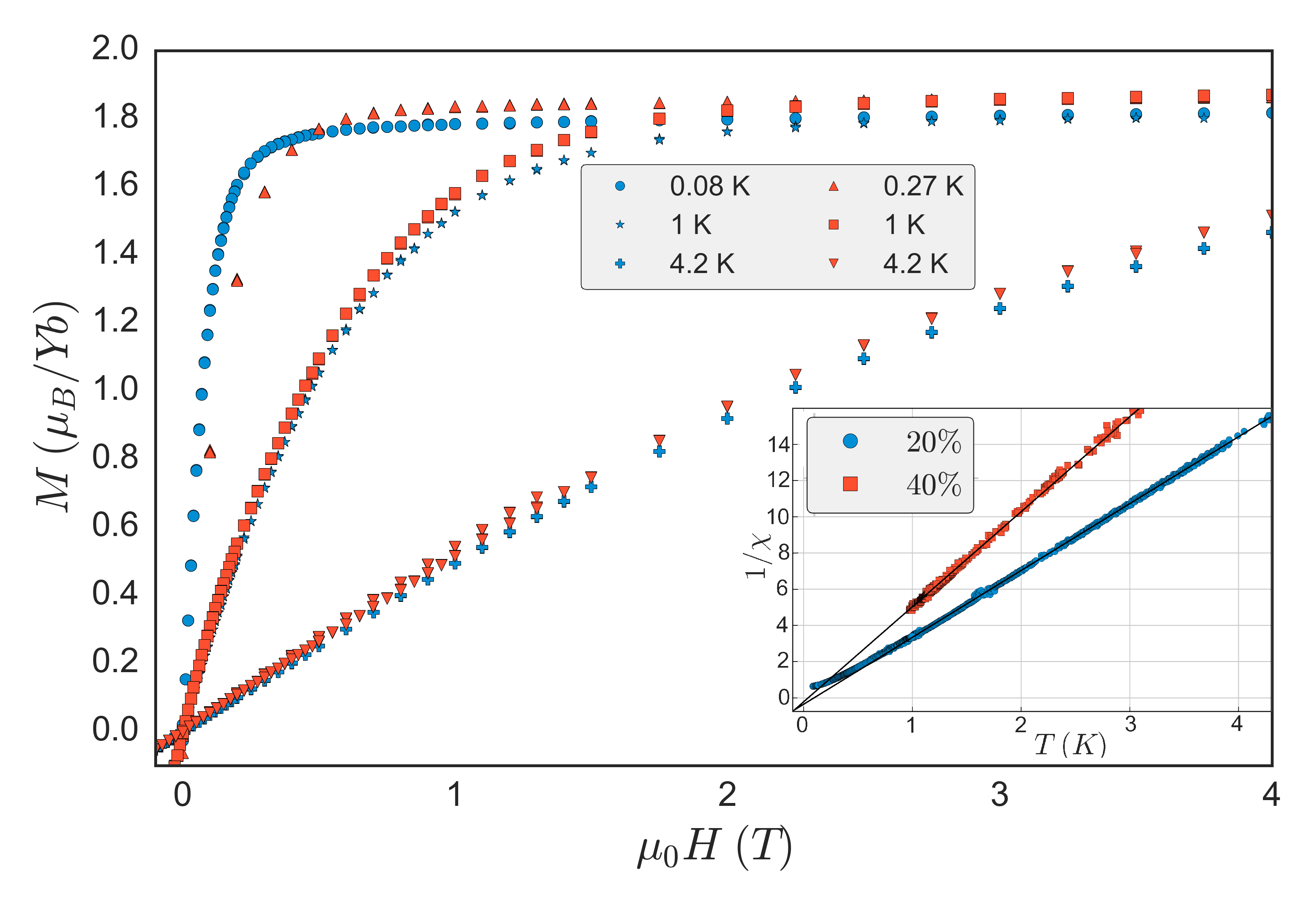}
\caption{Field dependence of the magnetization  for  diluted YbGG at several temperatures. Blue symbols show the experimental data for the 20\% diluted sample, red symbols represent the 40\% diluted sample.
Inset: Inverse volumic susceptibility as a function of temperature for the two diluted samples with fits to a Curie-Weiss law (black lines). The 20\% diluted data is in a field of 0.1~T, the 40\% sample data contain results for 0.1~T, 0.2~T and 0.4~T fields, which are all in the linear regime.
}
\label{MvsHdoped}
\end{figure}

We first focus on the low temperature magnetic properties of the substituted samples. 
For the 40\% diluted sample at 80~mK the magnetization (figure~\ref{MvsHdoped}) reaches an apparent saturation of 1.80 $\mu_{\rm B}$ per magnetic ion at 1~T. The magnetization can be increased slowly with increasing field up to a value of  1.86 $\mu_{\rm B}$ by 8~T, indicating the system is not truly saturated, as was observed in the pure sample \cite{Lhotel2021}. 
In the 20\% sample at 0.27~K a larger apparent saturated moment is observed of 1.85~$\mu_{\rm B}$ at 1~T. Similarly to the other sample the magnetization continues to increase slowly as the field is increased, reaching 1.9 $\mu_{\rm B}$ by 8~T. 
Both samples therefore exhibit saturated magnetic moments of  Yb$^{3+}$ ions that are slightly larger than for the pure sample. The difference reaches 5 \% in the 20\% sample. This could be due to an error in the estimation of the concentration of the Y$^{3+}$ ions, but the latter does not exceed 2 \%. Hence, the observed difference is  intrinsic and can come  from the small changes of the local crystal environment of the Yb$^{3+}$ ions.
The modified crystal electric field affects  the ground-state Kramers doublet producing a small change  in the saturated 
moments.


Low field magnetization measurements as a function of temperature were fitted to Curie-Weiss functions 
$\chi = C/(T-T_{\rm CW})$, assuming a linear behavior in an applied field ($\chi\approx M/H$) (inset of figure \ref{MvsHdoped}). 
In the 20\% sample, we find $T_{\rm CW}=92$ mK and $C=0.27$~K (corresponding to an effective moment of $\mu_{\rm eff}=3.14~\mu_{\rm B}$). In the 40\% sample, $T_{\rm CW}=42$ mK and $C=0.19$~K ($\mu_{\rm eff}=3.05~\mu_{\rm B}$). Regarding the effective moments, they are larger than in the pure sample ($\mu_{\rm eff}=3~\mu_{\rm B}$) \cite{Lhotel2021}, in a way that is consistent with the saturated magnetization measured in isothermal magnetization curves discussed above. 
The Curie-Weiss temperature in the 20\% sample is similar to the value in Ref. \onlinecite{Lhotel2021} ($T_{\rm CW}=97$~mK) for the pure sample, indicating that the dilution has not strongly affected the effective interactions. 
This certainly 
contradicts  to a picture of nearest-neighbor interactions in YbGG and indirectly supports the conclusion
of its dipolar nature. \cite{Lhotel2021}
Note however that an earlier study found $T_{\rm CW}=45$~mK \cite{Filippi1980}, which underlines the difficulty of determining precisely the Curie-Weiss temperature when it is that small, so that demagnetizing effects and Van Vleck susceptibility can alter the obtained value. In the 40\% sample, $T_{\rm CW}$ is reduced by a factor of about 2. This smaller $T_{\rm CW}$ (measured in similar conditions as the 20\% sample) suggests that the average interactions are smaller in the 40\% sample, as expected in the presence of dilution. Nevertheless, these results point out the robustness of the interactions in substituted YbGG, since the 20\% dilute sample is almost not affected, and ferromagnetic interactions are still present in the 40\% one. This may be due to the dipolar nature of the dominant interactions in these compounds. 

\begin{figure}[tb]
\centerline{
\includegraphics[width=0.8\linewidth,keepaspectratio]{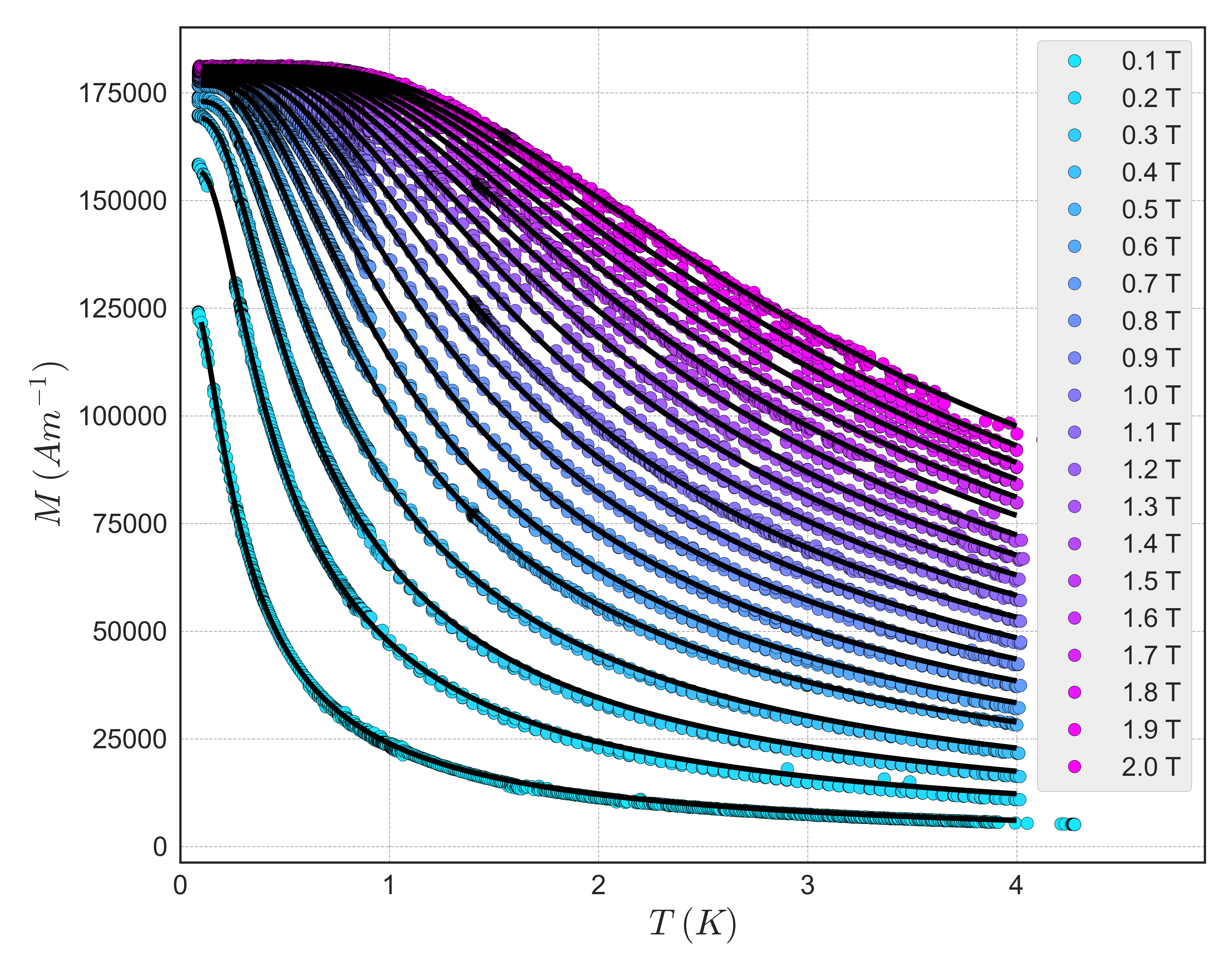}
}
\caption{Magnetization versus temperature for the 20\% diluted YbGG. Symbols represent 
the experimental data and black lines show the analytical fits.}
\label{MvsTdopeds}
\end{figure}

Turning now to the magnetocaloric properties, a
common figure of merit for  a refrigerant material is a field variation of  the magnetic entropy 
at a constant temperature $T$:
$\Delta S_m=S_m (T,B) - S_m (T,0)$.  By applying the Maxwell relations,
the entropy variation can be calculated as
\begin{equation}
\label{integration}
\Delta S_m (T,B) = \int_0^B \left( \frac{dM}{dT} \right)_{B'} dB' \ ,
\end{equation}
where the magnetization data for all intermediate fields $0\leq B'\leq B$ are used. In order to obtain reliable  
values of $\Delta S_m$, the integrand $dM/dT$ must be a smooth function of $B'$ and $T$. The raw experimental data $M(T,B)$ inevitably contain some noise. Therefore, we  fit $M(T,B)$
 normalized to its saturated value at $B\to\infty$ with
 a modified Brillouin function $B_{S}(T,B_{\rm eff})$, where $S=7/2$ is the spin of Gd$^{3+}$ ions
 and the effective magnetic field $B_{\rm eff}$ is taken as
\begin{equation}
 B_{\rm eff} = \sqrt{B^2 + b^2}  \ .
\end{equation}
Here the fit parameter $b$ stands for  an internal $T$-independent magnetic field, see, for example, Ref.~\onlinecite{Pobell1992}. 
The fits are only used to smooth the experimental data, hence, we do not discuss in the following the obtained values for $b$ 

Magnetization of the pure and substituted samples has been measured up to 4~T between 0.08~K and 4~K and is shown by the coloured points for the 20\% sample in figure~\ref{MvsTdopeds}. Only the data up to 2~T are shown since at low temperatures, the magnetization becomes almost saturated at larger applied fields. 
The modified Brillouin function form gives a good fit to the data across the entire temperature range and is therefore preferred to using data smoothing for finding the derivative. Taking the derivative analytically and performing the numerically integration in Eq.~\ref{integration} we obtain the  entropy variation curves  $\Delta S_m(T,B)$ shown in figure~\ref{entropydopeds}. 

\begin{figure}[t]
\centerline{\includegraphics[width=0.8\linewidth,keepaspectratio]{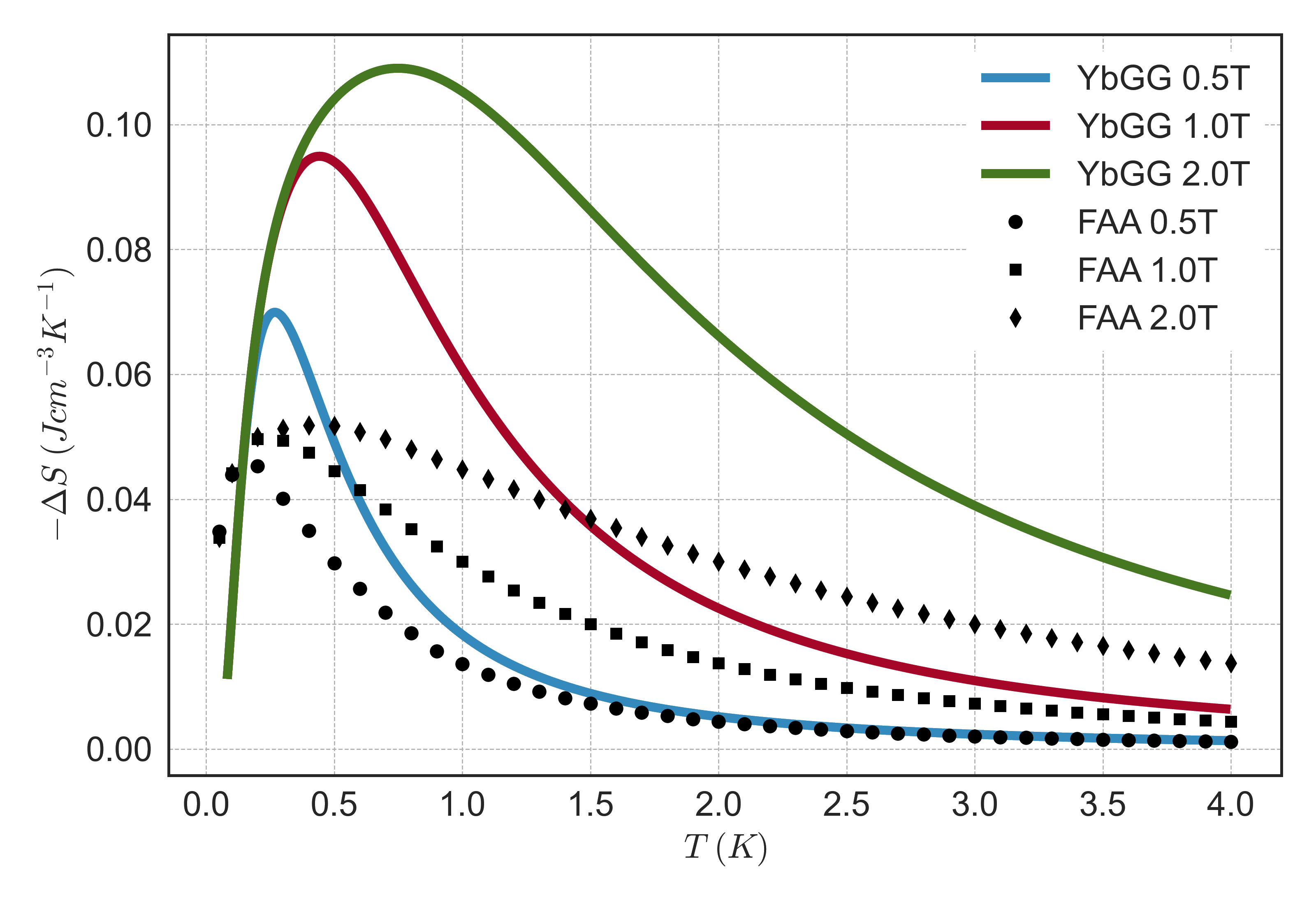}}
\vskip 3mm
\centerline{\includegraphics[width=0.8\linewidth,keepaspectratio]{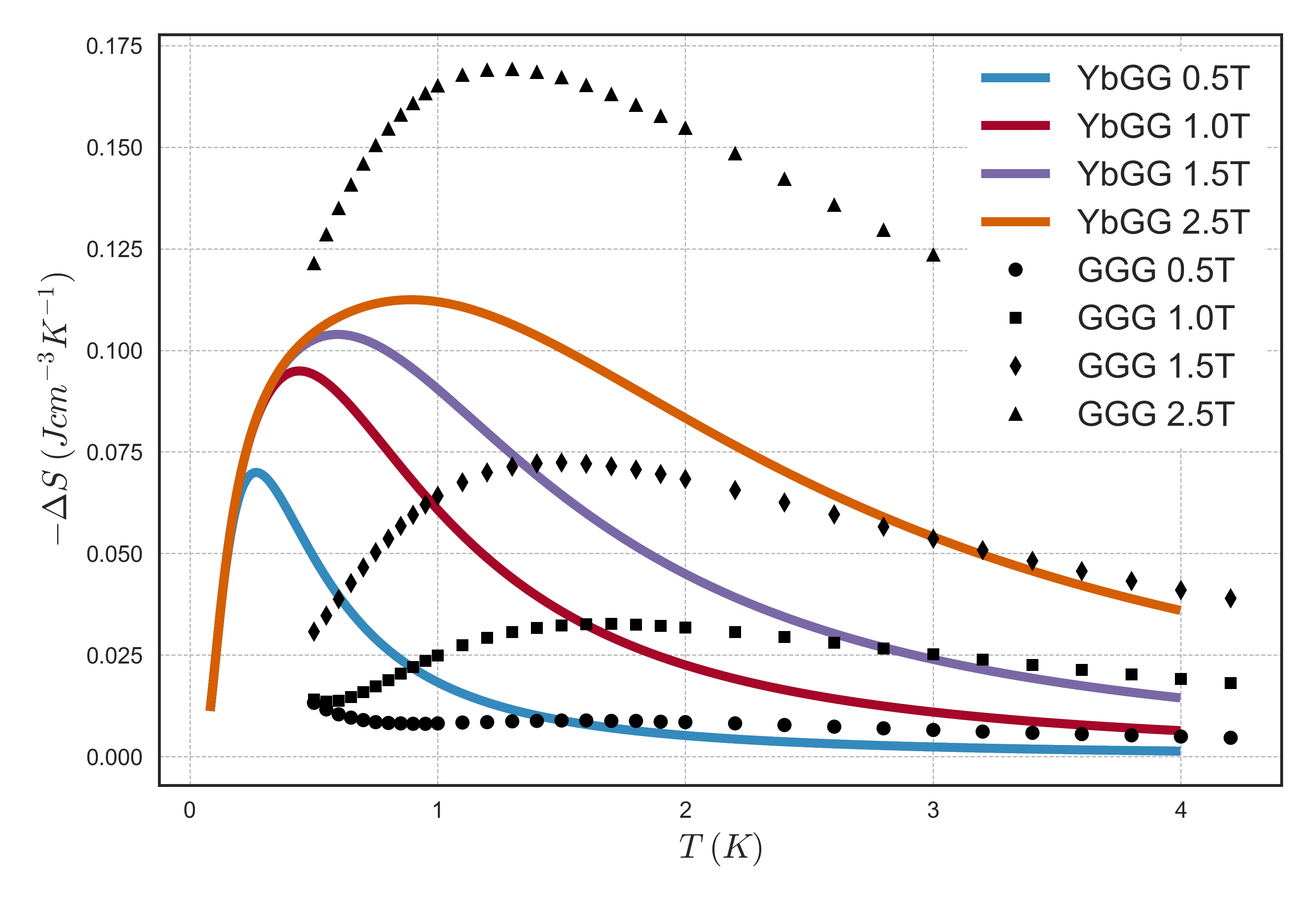}}
\caption{Isothermal entropy variations $\Delta S_m$ of  pure YbGG  compared to the reference 
magnetocaloric materials: FAA, \cite{Wikus2014} and GGG. \cite{Fisher1973} }
\label{comparisonS}
\end{figure}

In figure 3, the magnetocaloric properties of the pure compound for different applied fields are compared with other known magnetocaloric materials used at cryogenic temperatures, namely the paramagnetic salt FAA (ferric ammonium alum), \cite{Wikus2014} and the frustrated compound GGG (Gd$_3$Ga$_5$O$_{12}$). \cite{Fisher1973} This shows that YbGG  has very good performances compared to these materials, especially at low field. It is better than FAA at all fields up to 2 T between 200 mK and 4 K. Comparing with GGG it is significantly better in the very low temperature range, typically up to 1.5 K, for fields below 2 T. This ability to generate magnetocooling power at low field is interesting to limit the energy consumption associated to the applied magnetic field, which is especially relevant in the context of spatial applications.

\begin{figure}[tb]
\centerline{\includegraphics[width=0.8\linewidth,keepaspectratio]{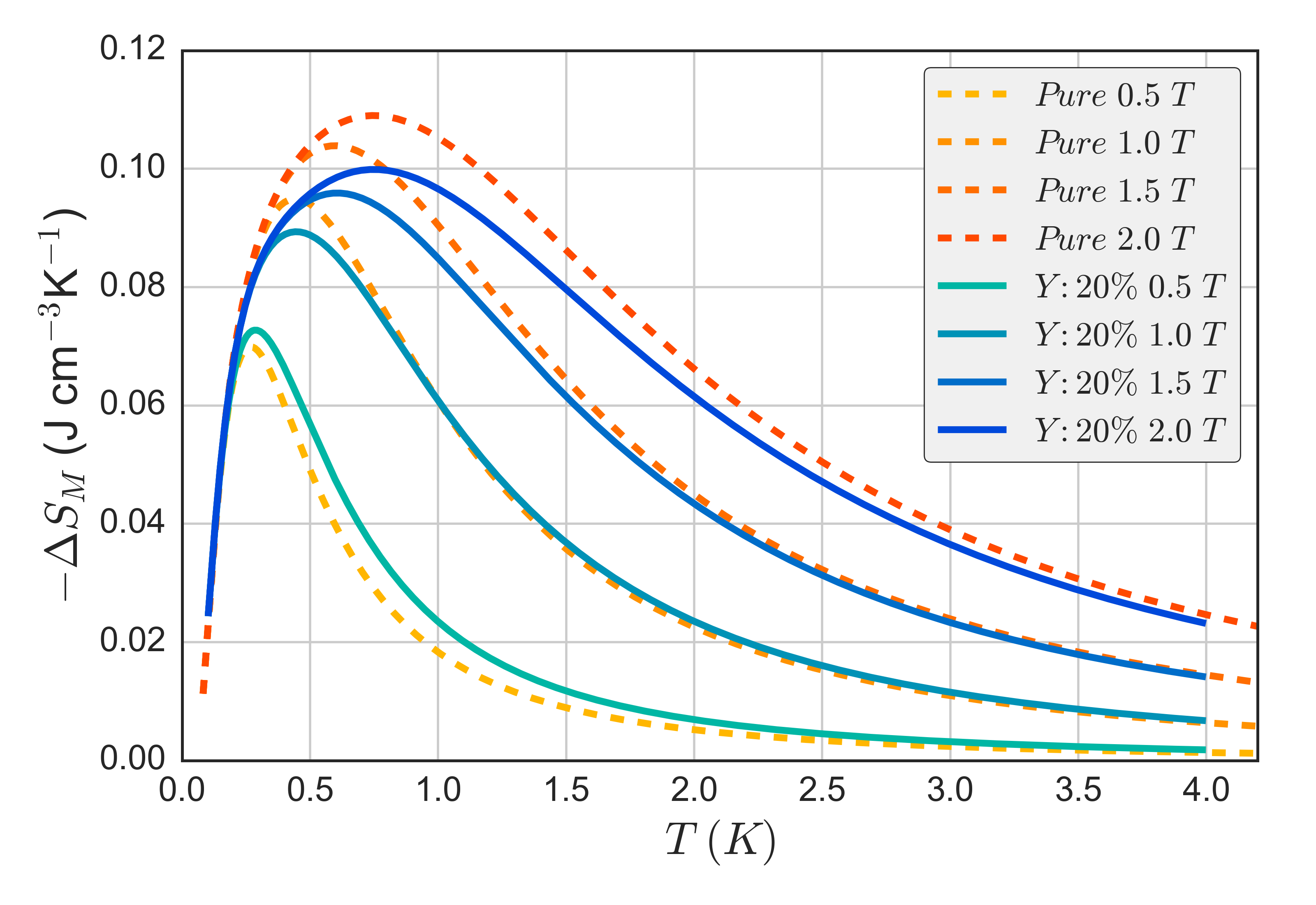}
}
\vskip 3mm

\centerline{\includegraphics[width=0.8\linewidth,keepaspectratio]{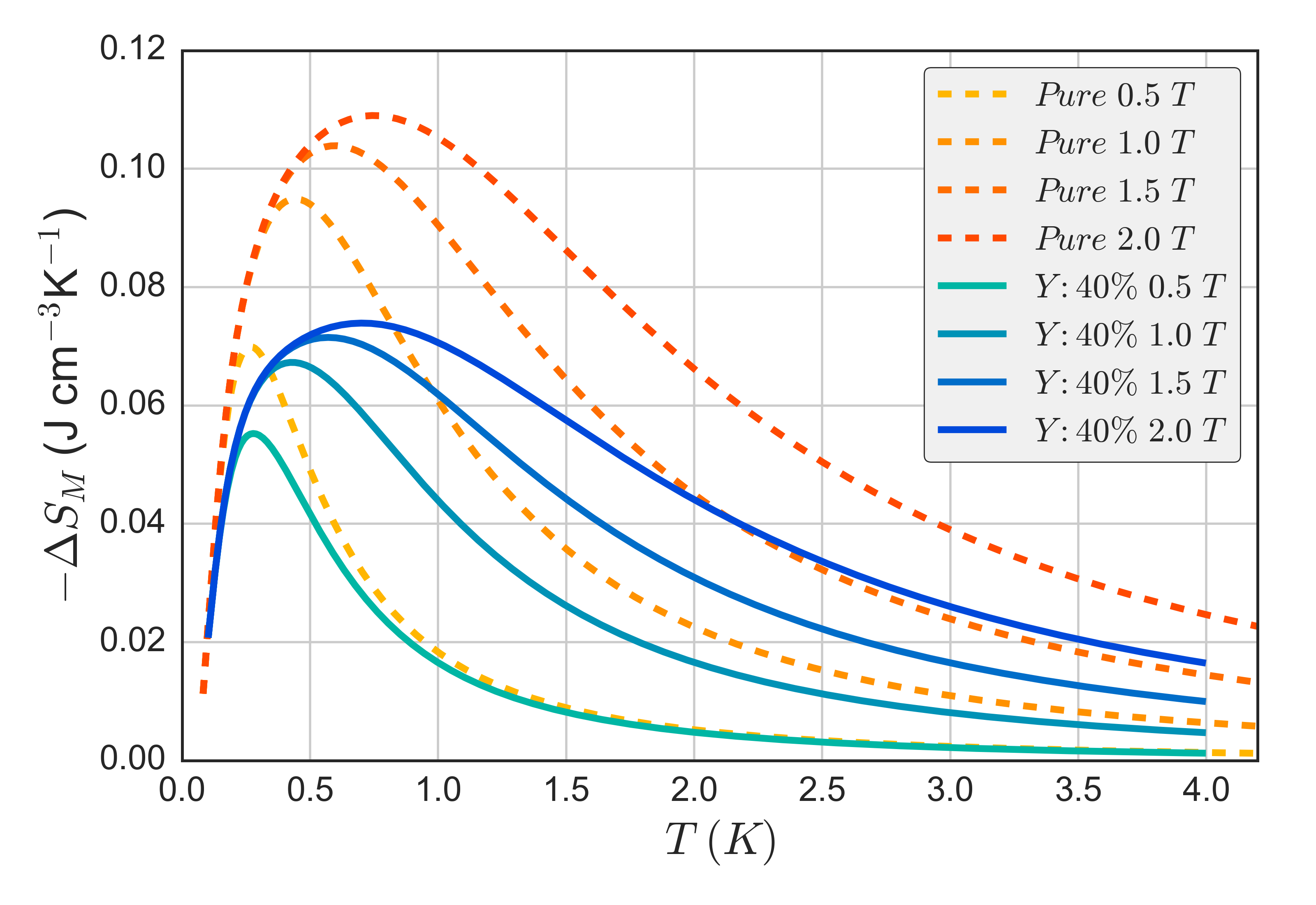}
}
\caption{Comparisons of $\Delta S_m$ between the pure sample (dashed lines in both) the 20\% sample (top figure dashed lines) and the 40\% sample (bottom figure dashed lines)}.
\label{entropydopeds}
\end{figure}

Our interest is to check the ability to keep these magnetocaloric properties when diluting the Yb magnetic moment, which could open new opportunities by modifying the mechanical properties or the thermal conductivity. 
In the 20\% substituted sample (figure \ref{entropydopeds} top), the entropy change for a given applied magnetic field is found to be similar to the pure sample. At low field (0.5 T), the entropy change is even marginally increased. When increasing the field above 1.5 T, the performances are slightly diminished, especially below 1.5 K.
Whilst is it important at this point to bear in mind that this data has not had demagnetization corrections applied (which tend to reduce the entropy change), even for the largest value of $D$ ($1/3$) this behaviour is not significantly diminished. 
The origin of this sustained magnetocaloric power is unknown but likely a result of a combination between the slightly larger moment of the Yb$^{3+}$ ion in the 20\% sample, and complex correlation effects that appear in the presence of the dilution. 
Further theoretical models including dilution should be helpful to understand thoroughly this result, and determine if an optimal concentration of Y can be achieved.

Our measurements find that $\Delta S_m$ in the 40\%sample (figure \ref{entropydopeds} bottom), demonstrates a smaller magnetocaloric effect than in the pure sample, with a maximum entropy change reduced by a factor of about 1.5. This means that the substitution here has played the role of a conventional dilution, despite the frustrated nature of the system, and thus has not reinforced the magnetocaloric effect.  


In summary, we have performed magnetic measurements on three samples, a pure single crystal sample and two powdered diluted samples with 20\% and 40\% substitution of Yb with Y ions. Measurements of the quasi-saturated magnetic moment have found larger than expected moments which is particularly true  for the 20\% sample. Curie-Weiss fits yield positive Curie temperatures indicating ferromagnetic correlations at low temperature. The magnitude of the magnetocaloric effect has been indicated in these samples by converting measurements of the magnetization to entropy change.  The volumetric entropy release for the  20\% diluted sample is in a slight excess to that for 
the pure sample. However, the magnetocaloric effect for the 40\% sample is further reduced in comparison  to the other two.
Since the entropy release depends non-trivially on the doping concentration,  it can be possible that another intermediate concentration of non-magnetic ions $0<x<0.2$ may yield a greater magnetocaloric effect normalized per unit volume.  Further investigation of the doping dependence presents an interesting avenue for future studies.
Finally, the persistence of a strong magnetocaloric effect in the presence of dilution is of strong interest for the application for magnetic cooling. Indeed, a difficulty arose in the applications from the poor thermal conductivity of these oxide materials at low temperature. Substituting YbGG with carefully chosen elements might provide a way of overcoming this difficulty, without altering the magnetocaloric properties.

\section*{Author declaration}

\subsection*{Conflict of interest}
The authors declare no conflict of interests.

\subsection*{Author contributions}
E. Lhotel and M. E. Zhitomirsky have planned and prepared the project work. 
N.-R. Camara and C. Marin have grown the samples. 
E. Riordan and E. Lhotel have performed magnetization measurements.
E. Riordan and M. E. Zhitomirsky have analyzed the experimental data.
E. Riordan, E. Lhotel and M. E. Zhitomirsky have written and prepared the manuscript.

\section*{Acknowledgements}
We are grateful to J.-P. Brison, J.-M. Duval, G. Knebel, and S. Raymond for collaboration on the related projects.
The work was partially supported by Agence Nationale de la Recherche, France, Grant No. ANR-18-CE05-0023.

\end{document}